\documentclass[12pt,final]{article}
\usepackage{a4wide,graphicx}

\begin{document}



\begin{titlepage}
\begin{flushright}
McGILL-03-02\\
UA/NPPS-02-03
\end{flushright}
\vspace{0.8in}

\begin{center}
{\bf \large Comparison of approximate to exact next-to-next-to leading order corrections for Higgs and pseudoscalar Higgs boson production}\\
\vspace{0.5in}
A.P.\ Contogouris$^{a,b,}$\footnote{e-mail: apcont@physics.mcgill.ca, acontog@cc.uoa.gr} and P.K. Papachristou$^{b,}$\footnote{e-mail: ppapachr@cc.uoa.gr}\\
\vspace{0.3in}
(a) \textit{Department of Physics, McGill University}\\
\textit{Montreal, Quebec, H3A 2T8, CANADA}\\
(b) \textit{Nuclear and Particle Physics, University of Athens}\\
\textit{Panepistimiopolis, Athens 15771, GREECE}
\end{center}
\begin{abstract}
Recently obtained NNLO exact corrections for Higgs and Pseudoscalar Higgs boson production in hadron colliders are compared with approximate ones. As shown before, it is found that there is a range of a proper variable where these corrections differ little.
\end{abstract}
\end{titlepage}

Some time ago it was argued that for processes involving structure functions and/or fragmentation functions, over a range of a proper kinematic variable $w$, there is a part that dominates the next- to leading order (NLO) correction and that this part contains the distributions $\delta(1-w)$ and $[\ln^n(1-w)/1-w)]_+$ $n=0,1$ \cite{CON90}. Subsequently this argument was extended to the then existing next- to- next- to leading order (NNLO) calculations, namely Drell- Yan (D-Y) production of lepton pairs ($q+\overline{q}\to\gamma^*$) and deep inelastic structure (DIS) functions ($q+\gamma^*\to q$) \cite{CON02}\footnote[1]{This paper, in preliminary form, was presented in "QCD 2000'', Montpellier, France, July 2000, and published in Nucl. Phys. B, (Proceed. Suppl.) \textbf{96} (2001) 94 (hep-ph/0109070).}.

In the meantime two more processes have been calculated in NNLO: Higgs boson production in hadron- hadron collisions ($g+g\to H$) \cite{ANA02} and neutral pseudoscalar Higgs boson production in hadron- hadron collisions ($g+g\to A$) \cite{ANA02b}. Clearly, it would be important to see whether the procedures developed in \cite{CON02} apply also to Higgs and pseudoscalar Higgs boson production as well.

In the calculation of Higgs boson production in hadron- hadron collisions to leading order (LO) \cite{ELL79} and to NLO \cite{DJO91} no approximations of the Higgs- two gluon vertex are necessary\footnote[2]{To NLO this is due to an accidental cancellation of the dependence on the top quark mass between real and virtual corrections. See the last paper of Ref. [6].}. This vertex is dominated by the top quark, which is known to have a mass $m_t$ much greater than that of the other quarks. However, the NNLO calculation was possible only in the limit of the Higgs mass $m_H$
\begin{equation}
m_H<<2m_t .
\end{equation}
In this limit the top-quark loops are replaced by point-like vertices and the corresponding effective Lagrangian \cite{DJO91,ELL76,VOL86} is known to provide a satisfactory description of the cross section for a Higgs boson at NLO \cite{DJO91}. 

In the calculation of the pseudoscalar Higgs boson production the situation is more complicated. The Higgs boson sector of the Minimal Supersymmetric Standard Model consists of two complex Higgs doublets. Thus, apart from the mass of the neutral pseudoscalar Higgs boson $m_A$, the ratio of the vacum expectation values of the two Higgs doublets $\upsilon_1/\upsilon_2\equiv\tan\beta$ also enters. The calculation of \cite{ANA02b} is valid for small and moderate values of $\tan\beta$; only then the $gg\to A$ is dominated by a top-quark loop. Then, for 
\begin{equation}
m_A<<2m_t
\end{equation}
the interaction of the pseudoscalar Higgs boson can be described by an effective Lagrangian \cite{CHE97}.

In our approach \cite{CON02} the proper variable is proportional to $\tau=m_H^2/S$ (or $\tau=m_A^2/S$), where $\sqrt{S}$ is the total c.m. energy of the initial hadrons. Our approach requires inclusion of the region $\tau$ large (inclusion of $\tau$ near 1). Since experiment excludes values of $m_H$ (or $m_A$) $\leq 100$ GeV, we have to consider $\sqrt{S}$ well exceeding this value. On the other hand, inclusion of $\sqrt{S}\geq 2$ TeV would require $m_H$ (or $m_A$) well exceeding 1 TeV, which would render questionable the field- theoretic approach. We then have considered a nominal energy of $\sqrt{S}=520$ GeV. Clearly, for $m_H$ (or $m_A$) $\geq 200$ GeV the inequality (1) (or (2) ) is violated and the whole results of \cite{ANA02} and \cite{ANA02b}, which we use, should be considered as just providing a mathematical model, where the approach of \cite{CON02} can be tested.
\\
\\
We begin with $pp\to H + X$ (or $p\overline{p}\to H+X$) mediated via the subprocess $gg\to H$\footnote[1]{We remind that in our approach \cite{CON90,CON02} the various perturbation orders (LO, NLO, NNLO) should refer to the same subprocess.} and we consider the cross- section
\begin{equation}
\sigma _{h_1  + h_2  \to H + X} (m_H^2 ,S) = \int\limits_0^1 {dx_1 dx_2 \overline f _{g/p} (x_1 )\overline f _{g/p} (x_2 )\sigma _{gg \to H} (m_H^2 ,x_1 x_2 S)} 
\end{equation}
where $h_1,h_2$ denote $p,p$ (or $p,\overline{p}$) and $\overline{f}_{g/p}(x)$ is the standard distribution of gluons inside the $p$ (or $\overline{p}$). Using dimensional analysis we write the partonic cross- section in terms of the dimensionless variable
\begin{equation}
z = \frac{{m_H^2 }}{{x_1 x_2 S}} = \frac{\tau }{{x_1 x_2 }}
\end{equation}
and after factoring the collinear singularities (usually in the $\overline{MS}$ scheme) we end up with the following expression \cite{ANA02}
\begin{equation}
\sigma _{h_1  + h_2  \to H + X} (\tau ,S) = \tau f_{g/p}  \otimes f_{g/p}  \otimes \left( {\sigma _{gg} (z)/z} \right)(\tau )
\end{equation}
where $\otimes$ denotes the standard convolution defined as
\begin{equation}
\left[ {f_1  \otimes f_2 } \right](\tau ) = \int\limits_0^1 {dx_1 dx_2 f_1 (x_1 )f_2 (x_2 )\delta (\tau  - x_1 x_2 )}.
\end{equation}
The partonic cross- section $\sigma_{gg}(z)$ is given by the following perturbation expansion
\begin{equation}
\sigma _{gg} (z) = \sigma _0 \left[ {\eta_{gg}^{(0)} (z) + \frac{{\alpha _s }}{\pi }\eta_{gg}^{(1)} (z) + \left( {\frac{{\alpha _s }}{\pi }} \right)^2 \eta_{gg}^{(2)} (z) + O(\alpha _s^3 )} \right]
\end{equation}
where the functions $\eta_{gg}^{(k)}$, $k=0,1,2$, are given in Eqs. (44), (45), (47), (48) and (49) of \cite{ANA02}, and
\begin{equation}
\sigma _0  = \frac{\pi }{{576\upsilon ^2 }}\left( {\frac{{\alpha _s }}{\pi }} \right)^2 
\end{equation}
with $\upsilon\approx 246$ GeV the Higgs vacum expectation value.
\\
\\
Subsequently we proceed as in \cite{CON02}\footnote[2]{Although known since long ago (see [1]), as in [2], we present also results for $k=1$.}. We write, for simplicity, $\sigma _{h_1  + h_2  \to H + X} (\tau ,S) \equiv \sigma _H (\tau ,S)$ and denote by $\sigma_{H}^{(k)}(\tau,S)$, $k=0,1,2$, the $O(\alpha_s^k)$ part of $\sigma _H (\tau ,S)$, by $\sigma_{Hs}^{(k)}$ the part of $\sigma_H^{(k)}$ arising from distributions $\delta(1-z)$ and $[\ln^n(1-z)/1-z)]_+$ , here $n=0,1,2,3$ (virtual, collinear and soft gluons) and by $\sigma_{Hh}^{(k)}$ the rest. We also define
\begin{equation}
L_H^{(k)} (\tau ,S) = \frac{{\sigma _{Hh}^{(k)} (\tau ,S)}}{{\sigma _H^{(k)} (\tau ,S)}}.
\end{equation}
In the subsequent calculations we use $n_f=5$ flavors and fix the renormalization and factorization scales at $\mu=M=m_H$. For the gluon distributions we use the updated $\overline{MS}$ CTEQ5M1 set of \cite{LAI00}. 

Fig. 1, upper part, shows $L_H^{(k)}$, $k=1,2$, as functions of $\sqrt{\tau}$. For $L_H^{(1)}$, while for relatively small $\sqrt{\tau}$ is significant, for $\sqrt{\tau}\geq .63$ is below 30\%. As for $L_H^{(2)}$, for $\sqrt{\tau}\geq .43$ is smaller than 20\%. Moreover, both $L_H^{(k)}$ decrease fast as $\sqrt{\tau}$ increases towards 1.

As in \cite{CON02}, it is of interest to see the precentage of $\sigma_{Hh}^{(k)}$ of the total cross section determined up to $O(\alpha_s^k)$. Fig. 1, upper part, also shows the ratios $\sigma _{Hh}^{(1)} /\left( {\sigma _H^{(0)}  + \sigma _H^{(1)} } \right)$ and $\sigma _{Hh}^{(2)} /\left( {\sigma _H^{(0)}  + \sigma _H^{(1)}  + \sigma _H^{(2)} } \right)$. The former is below 31\% and the latter below 15\% for all $\sqrt{\tau}$\footnote[1]{This can also be seen in the first paper of Ref.[11].}. Again, both ratios decrease fast as $\sqrt{\tau}$ increase.
\\
\\
Now we turn to the calculation of the pseudoscalar Higgs boson production and consider $pp\to A+X$ (or $p\overline{p}\to A+X$) mediated via the subprocess $gg\to A$. As before, the partonic cross-sections $\sigma_{gg}(z)$ have an expansion similar to (7)
\begin{equation}
\sigma _{gg} (z) = \sigma _0 \left[ {\phi _{gg}^{(0)} (z) + \frac{{\alpha _s }}{\pi }\phi _{gg}^{(1)} (z) + \left( {\frac{{\alpha _s }}{\pi }} \right)^2 \phi _{gg}^{(2)} (z) + O(\alpha _s^3 )} \right]
\end{equation}
where $z$ is given by (4) with $\tau=m_A^2/S$, $\phi_{gg}^{(k)}(z)$ are given in Eqs. (8)-(11) of \cite{ANA02b} (together with the expressions of $\eta_{gg}^{(k)}(z)$), and here
\begin{equation}
\sigma _0  = \frac{\pi }{{256\upsilon ^2 \tan ^2 \beta }}\left( {\frac{{\alpha _s }}{\pi }} \right)^2 .
\end{equation}
Now we write $\tan ^2 \beta \sigma _{h_1  + h_2  \to A + X}  \equiv \sigma _A (\tau ,S)$ and, as before, denote by $\sigma_A^{(k)}(\tau,S)$ the $O(\alpha_s^k)$ part of $\sigma_A(\tau,S)$, by $\sigma_{As}^{(k)}$ the part of $\sigma_A^{(k)}$ arising from distributions and by $\sigma_{Ah}^{(k)}$ the rest. We define
\begin{equation}
L_A^{(k)} (\tau ,S) = \frac{{\sigma _{Ah}^{(k)} (\tau ,S)}}{{\sigma _A (\tau ,S)}}
\end{equation}
and fix the renormalization and factorization scales at $\mu=M=m_A$. Again, for the gluon distributions we use the set CTEQ5M1 set of \cite{LAI00}.

Fig. 1, lower part, shows $L_A^{(k)}$, $k=1,2$, as functions of $\sqrt{\tau}$. All the results are similar as for $L_H^{(k)}$. Similar are also the results for the ratios $\sigma _{Ah}^{(1)} /\left( {\sigma _A^{(0)}  + \sigma _A^{(1)} } \right)$ and $\sigma _{Ah}^{(2)} /\left( {\sigma _A^{(0)}  + \sigma _A^{(1)}  + \sigma _A^{(2)} } \right)$.
\\
\\
We note the following\footnote[1]{A similar remark regarding resummations was first made by M. Kramer, E. Laenen and M. Spira, Nucl. Phys. B \textbf{511} (1998) 523.}: suppose that in $\sigma_{Hs}^{(k)}$ and $\sigma_{As}^{(k)}$, apart from the terms arising from the distributions $\delta(1-z)$ and $[\ln^n(1-z)/1-z)]_+$ we include also the terms $\ln^{m}(1-z)$, $m=1,2,3$. Defining as $\sigma_{Hs}^{(k)}$ and $\sigma_{Ah}^{(k)}$ the rest, we find that the ratios $\sigma_{Hh}^{(k)}/\sigma_H^{(k)}$ and $\sigma_{Ah}^{(k)}/\sigma_A^{(k)}$ decrease significantly in magnitude over the entire range of $\sqrt{\tau}$. Of course, the same holds for the ratios $\sigma _{Hh}^{(1)} /\left( {\sigma _H^{(0)}  + \sigma _H^{(1)} } \right)$, $\sigma _{Hh}^{(2)} /\left( {\sigma _H^{(0)}  + \sigma _H^{(1)}  + \sigma _H^{(2)} } \right)$ and the corresponding ratios with H replaced by A.
\\
Note also that Ref.\cite{HAR02} has obtained numerical results very similar to \cite{ANA02} and \cite{ANA02b} by expanding the phase-space integrals around the kinematic point $z=\tau/x_1x_2=1$, where $\tau=m_H^2/S$ or $\tau=m_A^2/S$, and keeping a number of terms. Although the first paper of [11] was published before [3], we prefer the methods of [3] as they avoid expansions.\footnote[2]{In the first of Ref.\cite{HAR02} an error was found in the calculation of T. Matsuura et al., Nucl. Phys. B \textbf{319} (1989) 570 on D-Y production. We have repeated the relevant calculations of \cite{CON02} and found no significant change.}
\\
\\
Finally, in Fig. 2, upper part, we present the total cross-sections $\sigma _H^{(0)}  + \sigma _H^{(1)}  + \sigma _H^{(2)}$ (dashed line) and $\sigma _H^{(0)}  + \sigma _{Hs}^{(1)}  + \sigma _{Hs}^{(2)}$ (solid line). What is important is that as $\sqrt{\tau}$ increases towards 1 both cross-sections approach each other , and for $\tau\ge 0.8$ practically coincide. The same is observed in Fig. 2, lower part, which shows the quantities $\sigma _A^{(0)}  + \sigma _A^{(1)}  + \sigma _A^{(2)}$ (dashed) and $\sigma _{A}^{(0)}  + \sigma _{As}^{(1)}  + \sigma _{As}^{(2)}$ (solid).
\\
\\
In conclusion, under the assumptions discussed at the beginning, we have shown that not only in D-Y production and DIS \cite{CON02}, but also in Higgs and pseudoscalar Higgs boson production ($gg\to H$ and $gg\to A$) there is a part containing the distributions $\delta(1-z)$ and $[\ln^n(1-z)/1-z)]_+$, here $n=0,1,2,3$ (virtual, soft and collinear part) that for $\sqrt{\tau}$($=m_H/\sqrt{S}$ or $m_A/\sqrt{S}$) not too small dominates the NLO and NNLO correction. This part is determined much easier than the NLO and in particular the NNLO correction. Of course, as it was stressed in \cite{CON02}, this part should not be restricted to too small a region near 1, for threshhold resummation \cite{KID01} becomes very important.
\\
\\
NOTE ADDED\\
After the completion of this article, the paper by V. Ravindran, J. Smith and W. van Neerven, hep-ph/0302335, appeared, confirming the results of [3], [4] and [11] by a different method.

\section*{Acknowledgments}
We are much indebted to C. Anastasiou and K. Melnikov for very useful private communications and correspondence. A correspondence by R. Harlander and W. Kilgore is also greatfully acknowledged. The work was also supported by the Natural Sciences and Engineering Research Council of Canada, by the Research Committee of the University of Athens and by the Greek State Scholarships foundation (IKY).

\newpage

\begin{figure}[h!]
\centering
\includegraphics[height=18cm]{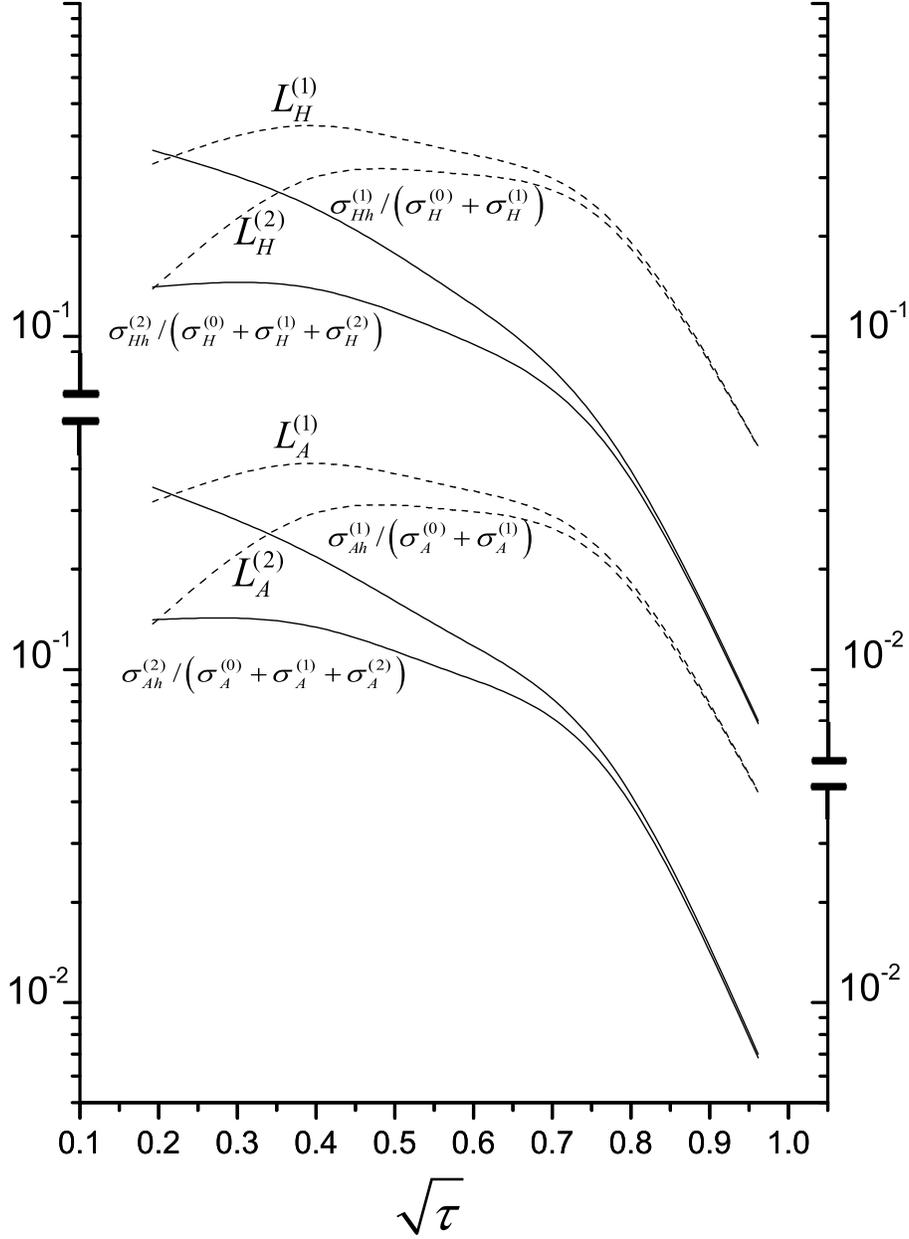}
\caption{Upper part: The ratios $L_H^{(1)}$ and $\sigma _{Hh}^{(1)} /\left( {\sigma _H^{(0)}  + \sigma _H^{(1)} } \right)$ (dashed lines) and the ratios $L_H^{(2)}$ and $\sigma _{Hh}^{(2)} /\left( {\sigma _H^{(0)}  + \sigma _H^{(1)}  + \sigma _H^{(2)} } \right)$ (solid lines) versus $\sqrt{\tau}=m_H/\sqrt{S}$. Lower part: The quantities $L_A^{(1)}$ and $\sigma _{Ah}^{(1)} /\left( {\sigma _A^{(0)}  + \sigma _A^{(1)} } \right)$ (dashed) and the quantities $L_A^{(2)}$ and $\sigma _{Ah}^{(2)} /\left( {\sigma _A^{(0)}  + \sigma _A^{(1)}  + \sigma _A^{(2)} } \right)$ (solid) versus $\sqrt{\tau}=m_A/\sqrt{S}$. }
\end{figure}

\begin{figure}[h!]
\centering
\includegraphics[height=18cm]{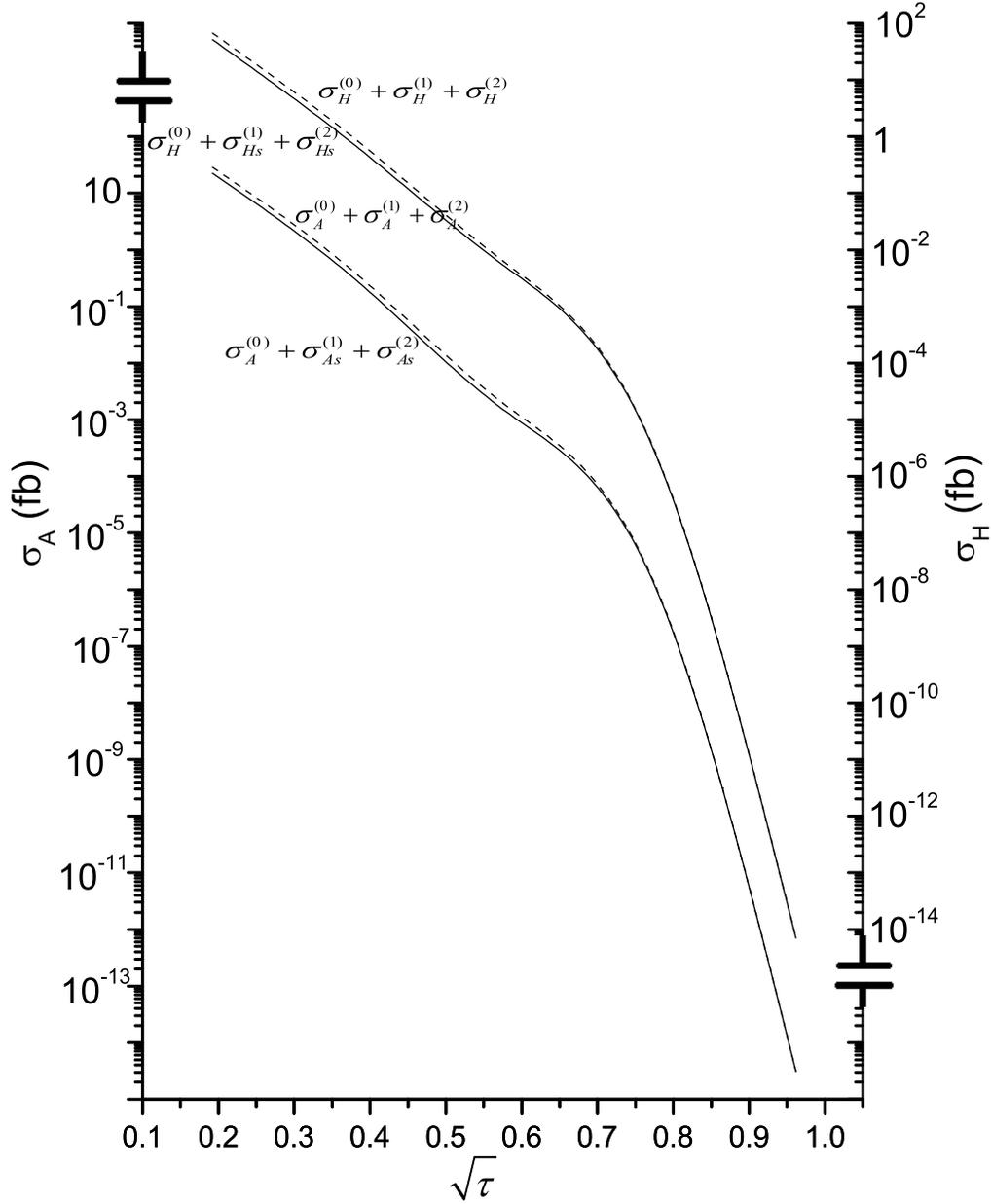}
\caption{ Upper part: The cross-sections $\sigma _H^{(0)}  + \sigma _H^{(1)}  + \sigma _H^{(2)}$ (dashed line) and $\sigma _H^{(0)}  + \sigma _{Hs}^{(1)}  + \sigma _{Hs}^{(2)}$ (solid line) versus $\sqrt{\tau}=m_H/\sqrt{S}$. Lower part: The quantities $\sigma _A^{(0)}  + \sigma _A^{(1)}  + \sigma _A^{(2)}$ (dashed) and $\sigma _A^{(0)}  + \sigma _{As}^{(1)}  + \sigma _{As}^{(2)}$ (solid) versus $\sqrt{\tau}=m_A/\sqrt{S}$.}
\end{figure}
\end{document}